# Ion irradiation of Fe-Fe oxide core-shell nanocluster films: Effect of interface on stability of magnetic properties

John S. McCloy[1], Weilin Jiang, Timothy C. Droubay, Tamas Varga, Libor Kovarik

*Pacific Northwest National Laboratory, 902 Battelle Blvd., PO Box 999, Richland, WA 99352, USA*

Jennifer A. Sundararajan, Maninder Kaur, You Qiang

*Department of Physics, University of Idaho, Moscow, ID 83844, USA*

Edward C. Burks, Kai Liu

*Department of Physics, University of California, Davis, CA 95616, USA*

**Abstract**

A cluster deposition method was used to produce films of loosely aggregated nanoclusters (NC) of Fe core-$Fe_3O_4$ shell or fully oxidized $Fe_3O_4$. Films of these NC on Si(100) or MgO(100)/$Fe_3O_4$(100) were irradiated to $10^{16}$ $Si^{2+}$/$cm^2$ near room temperature using an ion accelerator. Ion irradiation creates structural change in the NC film with corresponding chemical and magnetic changes which depend on the initial oxidation state of the cluster. Films were characterized using magnetometry (hysteresis, first order reversal curves), microscopy (transmission electron, helium ion), and x-ray diffraction. In all cases, the particle sizes increased due to ion irradiation, and when a core of Fe is present, irradiation reduces the oxide shells to lower valent Fe species. These results show that ion irradiated behavior of the nanocluster films depends strongly on the initial nanostructure and chemistry, but in general saturation magnetization decreases slightly.

---

[1] Current affiliation School of Mechanical & Materials Engineering, Washington State University, Pullman, WA; john.mccloy@wsu.edu

1.  INTRODUCTION

Magnetic nanomaterials and nanostructures have gained popularity in recent years because of their interesting properties and promising applications in various fields.[1-6] Bulk synthesis of such nanomaterials for commercial use and bringing controlled property changes in nanomaterials has been a great challenge for nanotechnology industries for more than two decades. Core-shell magnetic structures have been particularly interesting for researchers searching for a wide range of applications ranging from magnetic recording[7] to microwave absorption[8] to cancer treatments[9] and medical imaging.[10-12]

Particle-irradiation on nanomaterials has been suggested as a method to bring controlled property changes in nanomaterials.[13-15] The prominent behaviors observed in nanomaterials due to irradiation effects are the change in structural and magnetic properties. Numerous investigations in the past have exposed the extrinsic impacts of ion-irradiation in nanomaterials, including thin films,[16,17] nanoparticles,[18,19] nanoparticle embedded matrices,[20,21] nanotubes,[22] nanowires,[23] and nanoparticle-granular films, where the irradiation either enhances or degrades the structural and magnetic properties. The possible role for irradiation in fabrication of next-generation devices calls for the understanding of irradiation impacts in nanomaterials. Researchers in this field have studied the property changes, including the variation in structural and magnetic properties by varying the ion-irradiation dose,[24] incident energies of ions,[25] and duration of exposure.[26] Most of the literature on ion irradiation effects on magnetic nanoparticle systems has focused on solely metal systems. In previous work, we have discussed the effects of ion irradiation on $FeO$-$Fe_3N$,[27] $Fe_3O_4$,[28] and $Fe/Fe_3O_4$[29] nanocluster films. In this work, we extend the investigation of ion irradiated core-shell nanocluster films and consider the phase evolution, the oxidation state of the iron as a function of position in the structure, and the resulting magnetic properties.

## 2. EXPERIMENTAL

### a. Nanocluster synthesis

The iron-iron oxide nanoclusters themselves were synthesized using magnetron sputtering combined with gas aggregation method as described in detail previously.[30] The cluster size formed in the aggregation chamber was controlled by adjusting the aggregation distance, sputter power, He to Ar gas flow rate, and temperature inside the aggregation chamber. The Fe atoms were sputtered from a Fe target placed on the magnetron gun in the aggregation chamber by supplying 350 standard cubic centimeters per minute (sccm) of Ar gas and 50 sccm of He gas. These Fe atoms were allowed to react with 3 sccm oxygen by supplying the gases into the reaction chamber, or inside the aggregation chamber, to form the desired nanoparticle. These particles were subsequently deposited onto a substrate, Si(100) or $Fe_3O_4$ (100) film on MgO(100), in the NC deposition chamber to form the granular films. The majority of the characterization of the samples deposited on Si has been reported previously, but some data is included here for comparison; the samples deposited on MgO have not been reported previously. A schematic of the cluster deposition system is shown in Figure 1, and a summary of the deposition parameters for the NC films is shown in Table 1. For comparison, two sizes of core-shell samples previously studied[31] are also listed.

### b. Characterization

Grazing-angle incidence x-ray diffraction (GIXRD), a technique which eliminates the strong diffraction peaks from the single-crystal substrate and underlying $Fe_3O_4$ film, in the case of MgO substrates, was employed to study the crystallographic phase and average size of the NC crystalline grains at room temperature, both before and after ion irradiation. A Philips X'pert Multi-Purpose Diffractometer (MPD, PANalytical, Almelo, The Netherlands) with Cu Kα radiation ($\lambda$ = 0.154 187 nm) at a 5° incident angle was used for performing GIXRD. Phase quantification was performed using the collected GIXRD spectra and Topas (Bruker AXS Topas® 4.2) whole pattern fitting.

NC film microstructures and thicknesses were examined using a helium ion microscope (HIM, Orion Plus, Carl Zeiss SMT, Peabody, MA), at 30 keV and 0.1-0.4 pA, before and after $Si^{2+}$ ion irradiation. Transmission electron microscopy (TEM) was performed with an FEI Titan 80-300 equipped with a CEOS Cs –image aberration corrector and operated at 300 kV. The bright-field imaging (TEM-BF) was utilized to study the morphology of the particles. In addition, high resolution TEM was employed to study crystallographic nature of individual particles at the atomic scale. All images were recorded with Gatan US 1000 CCD camera and analyzed with Digital Micrograph 1.95. Sample preparation for TEM involved detaching of the sputtered $Fe/Fe_xO_y$ particles from the substrate with a razor blade, and then subsequent dry transfer on the lacey carbon-coated Cu grids. The Cu Grids were then directly loaded to a conventional double tilt TEM holder and transferred to the TEM.

Room temperature magnetic properties of the samples were studied with a vibrating sample magnetometer (VSM, Lakeshore Cryotronics, Westerville, OH). Hysteresis curves were collected up to 10 kOe in 250 Oe steps. A subset of films were also studied using first-order reversal curves (FORC)[32-34] using a MicroMag 2900 vibrating sample magnetometer (Princeton Measurements Corp., Princeton, NJ). FORC measurement conditions for the various samples were as follows: (S1): 286 FORCs, 40 Oe steps up to 8.5 kOe with averaging time of 100 ms; (S2): 286 FORCs were taken with 18 Oe steps up to 3 kOe with averaging time of 100 ms; (#2, #4): 120 FORCs were taken with 69 Oe steps up to 6.5 kOe with averaging time of 1 s. These settings were chosen to accommodate reasonable measurement time with sufficient resolution based on the hysteresis loop for each sample. To investigate magnetic characteristics in the samples, measured magnetization as a function of applied field ($H$) and reversal field ($H_r$) was used to compute the FORC distribution:[32,35]

$$\rho(H, H_r) = \frac{\partial^2 M(H, H_r)}{\partial H \partial H_r}. \tag{1}$$

The FORC diagrams were then plotted following the standard coordinate transformation to bias field, $H_b = (H + H_r)/2$, and coercive field, $H_c = (H - H_r)/2$. Since measurements were taken at room temperature, the bias axes represent interparticle dipolar interactions.[36,37]

### c. Nanoclusters on magnetite films

To test the effects of pre-existing $Fe_3O_4$ (magnetite) films on NC deposition, additional samples were prepared. MgO (100) (9.35 mm diameter circular or 10 mm square) substrates were used to deposit epitaxial films of $Fe_3O_4$ by pulsed laser deposition (PLD).[38] To minimize the deposition of molten droplets and particles during the laser ablation process, an off-axis growth configuration was utilized. All growths were performed in 10 mTorr of $N_2$. Pressure was controlled using a combination of a mass flow controller and an automatically-throttled gate valve between the chamber and the primary system turbopump. A KrF laser (248 nm, 2.4 J cm$^{-2}$ and 1-20 Hz) was rastered across a rotating target. The substrate holder was also rotating which, along with laser rastering and target rotation, provided uniform deposition across the substrate. Both thick and thin films of PLD $Fe_3O_4$ were grown on MgO at 350 °C to assess effects on subsequent NC growth. PLD films were measured by GIXRD, regular $\theta$-$2\theta$ scan, and x-ray reflectivity (XRR) and determined to be epitaxial single crystals, (100) $Fe_3O_4$ on (100) periclase MgO, with thickness of 4.4 - 4.5 nm (XRR, thin films) or 120 nm (Helium ion microscopy, thick film).

Prior to sending samples for NC deposition, major hysteresis loops were obtained by VSM for one MgO/$Fe_3O_4$(thin) and one MgO/$Fe_3O_4$(thick), and films were left in remnant state after saturation when NC were subsequently deposited. An additional sample of MgO/$Fe_3O_4$(thin) was supplied for NC deposition without magnetization. Thus the three samples are denoted M1-D (thin PLD, demagnetized state upon NC deposition), M1-R (thin PLD, remnant state upon NC deposition), and M2 (thick PLD, remnant state upon NC deposition). It was hypothesized that the deposition of NC on the remnant versus demagnetized PLD films might influence their aggregation or alignment.

### d. Ion irradiation

Ion irradiation of the NC samples was performed using a 3.0 MV electrostatic tandem accelerator (NEC 9SDH-2 pelletron, Middleton, WI). Each granular film was irradiated at normal incidence with 5.5 MeV $Si^{2+}$ ions to a fluence of $10^{16}$ ions/cm$^2$ near room temperature. A beam rastering ensured uniform irradiation over an area covering the entire sample surface. Typical ion flux was on the order of 0.01 ($Si^{2+}$

/nm$^2$)/s, and resulting increase of sample temperature was less than 50 K during irradiation. Computer simulations with the Stopping and Range of Ions in Matter (SRIM) code[39] indicates that the Si$^{2+}$ ions were implanted near the film-substrate interface, within the substrate and not within the NC films.

## 3. RESULTS AND DISCUSSION

### a. Structural Characterization

All nanocluster films were analyzed using GIXRD. Sizes are estimated using the line-broadening and the Scherrer equation, as previously described for these systems.[28] A summary of the constituent phases, their weight ratio and the average crystallite sizes of previously studied films and those in this study is shown in Table 2. The single phase Fe$_3$O$_4$ NC film on Si, described previously, consists of 3 nm primary particles before irradiation (S1-u) and 23 nm particles after irradiation (S1-i).[28] Before irradiation, the two-phase core-shell NC on Si has 8 nm Fe cores (S2-u) based on the well-resolved Fe(110) peak, and 2 nm Fe$_3$O$_4$ shells, based on the Fe$_3$O$_4$(311).[29] These values are typical and have been confirmed on similar samples by transmission electron microscopy (TEM).[36] In this sample after irradiation (S2-i), the Fe core shrinks slightly to 7 nm but the shell transformed to FeO with crystallite size 13 nm, as determined by whole pattern fitting. As seen by TEM (Figure 4, discussed below), this crystallite size is no longer just the shell thickness but represents the FeO matrix between two Fe cores. Overall, this represents a film content of ~81 wt% Fe and 19 wt% Fe$_3$O$_4$ before irradiation and ~36 wt% Fe and ~64 wt% FeO after irradiation.

For the iron-iron oxide NC films grown on MgO(100)/Fe$_3$O$_4$(100), there was little difference among the three samples in this study: 1) NC deposited onto a thin Fe$_3$O$_4$ in a demagnetized state: M1-D; 2) NC deposited onto a thin Fe$_3$O$_4$ in a remnant state: M1-R; and 3) NC deposited onto a thick Fe$_3$O$_4$ in a remnant state: M2. Before irradiation, films (M1-D-u, M1-R-u, M2-u) were largely 2 nm crystallite Fe$_3$O$_4$ (~77 wt%), a small amount of ~9 nm Fe cores (~2 wt%), and a substantial amount of ~8 nm crystallite orthorhombic (*Cmcm*) Fe$_2$O$_3$ (~21 wt%) (see Figure 2). There was no structure file available for fitting the orthorhombic $Fe_2O_3$ phase, so a similar orthorhombic $GeMgO_3$ phase was used for fitting the diffraction pattern. Recently, Machala[40] has reviewed the forms of $Fe_2O_3$, including $\alpha$-$Fe_2O_3$ (hematite), $\beta$-$Fe_2O_3$, $\gamma$-$Fe_2O_3$ (maghemite), $\delta$-$Fe_2O_3$, and the various high pressure forms of $Fe_2O_3$. One high pressure form of $Fe_2O_3$ was first identified by Ono et al.[41] at pressures >50 GPa at room temperature and described as "post-perovskite." This phase was later refined to be the $CaIrO_3$-type structure with orthorhombic space group *Cmcm*, and found to be stable from 68 GPa at 1200 K up to 96 GPa at 2300 K, the highest pressures and temperatures tested.[42] It is possible that this phase is formed metastably as a surface phase in the nanocluster system. It has been shown that defect phases of iron oxide as well as thermodynamically unexpected FeO can form as "buffer layers" in the nucleation of reduction of hematite or oxidation of magnetite.[43]

After irradiation, the $NC/Fe_3O_4/MgO$ films showed again similar phases to each other but drastically different from the unirradiated films (see Figure 2). First, the irradiated NC samples (M1-D-i, M1-R-i, M2-i) still consisted of mostly $Fe_3O_4$ (84-90 wt%) but with 15-18 nm crystallite size, which is a an increase in overall NC fraction and crystallite size from the unirradiated (77 wt%, 2 nm $Fe_3O_4$). Second, the irradiated samples also contain zero valent Fe (5 – 8 wt%) with crystallite sizes 29-31 nm. This represents both an increase in fraction and increase in size of the Fe portion from the unirradiated (2 wt%, 9 nm Fe). Third, the samples contain a fraction of wustite FeO (4-9 wt%) with crystallites 24-26 nm in size, a phase not present in the unirradiated samples. All the observed crystalline phases before and after ion irradiation of the films are summarized in Table 3. Note that there exists a crystalline phase of $(MgO)_{0.593}(FeO)_{0.407}$ (PDF#: 01-077-2367) that is indistinguishable from FeO (PDF#: 04-006-5424). It is possible that the $Si^{2+}$ ion irradiation at the iron-oxide/MgO interface induces atomic intermixing and promotes formation of the new phase. However, since FeO was observed also in the irradiated sample grown on Si (S2-i), it is likely that the FeO phase in the samples with MgO is due at least in part to the NC film and not merely changes in the interface between the PLD film and the MgO substrate. The increase in crystallite size of both the $Fe_3O_4$ and Fe components of the NC suggest radiation-induced grain growth, similar to the process observed in ion-irradiated $ZrO_2$,[44-46] $SnO_2$[47] and SiC.[48] It was also observed[26] under TEM that electron-irradiation could also induce thickening of the oxide layers on the surface of Fe nanoparticles. A decrease in oxidation state of the iron in the NC after irradiation is suggested for both the NC/$Fe_3O_4$/MgO samples, where $Fe_3O_4$ content decreases and Fe and FeO content increases, and the NC/Si sample, where $Fe_3O_4$ content is eliminated in favor of FeO (see Table 2).

Figure 3 shows a typical HIM cross-section images of the films before irradiation. The unirradiated film exhibits loosely interconnected nanoparticles similar to those observed for other Fe/$Fe_3O_4$ core-shell[36] and $Fe_3O_4$[28] clusters. Particles are agglomerated at various scales before irradiation, and films are highly porous with large surface and interface areas. From Fig. 3, the apparent thickness of the granular film for sample M1-D is ~4.6 μm. Rutherford backscattering spectrometry (RBS) and accompanying simulation shown in Fig. 3b indicates that the film thickness is ~1.3 μm based on the theoretical density of $Fe_3O_4$ (5.2 g/cm$^3$), confirming that the granular film is highly porous. Irradiation leads to grain growth and particle aggregation as described previously for $Fe_3O_4$ nanoclusters,[28] resulting in a densified continuous film with an enhanced adhesion to the substrate.

Figure 4 shows TEM images of unirradiated (a,c) and irradiated (b,d) core-shell particles, which confirm the conclusions obtained from GIXRD pattern fitting and from HIM. Particle sizes have some distribution, as has been observed in previous studies,[36] but the increase in particle size with ion irradiation is apparent. Again in confirmation with previous studies[49,50] on cluster-deposited core-shell iron/iron oxides, the larger cores in the irradiated samples are faceted along particular crystallographic planes, while the smaller cores in unirradiated samples are more spherical. For particles larger than 10 nm, both core and shell appear polycrystalline, and contain zones which are highly disordered or amorphous. Due to the inhomogeneity of the samples, it is not possible to quantitatively assess the level of disorder and amorphous fraction change between unirradiated and irradiated samples by TEM.

Electron beam-induced thickening of the oxide shell in similar core-shell particles has been observed previously by TEM and attributed to beam-enhanced mass transport by oxygen vacancy creation,[26] but that work did not observe a simultaneous reduction of the iron species as we observe here with ion irradiation. Rather, in our study, the ion irradiation appears to increase the Fe core size, increase the $Fe_3O_4$ shell thickness, and result in a small fraction (<10 wt% by XRD for the M1 and M2 samples) of FeO which may be present as disordered phase in the oxide shell. For these samples (deposited on $MgO/Fe_3O_4$) the substantial amount of $Fe_2O_3+Fe_3O_4$ (~97 wt%) prior to irradiation resulted in irradiated samples with <10 wt% FeO, while the sample with only 20 wt% $Fe_3O_4$ (and no $Fe_2O_3$, sample S2-u) prior to irradiation resulted in a much more substantial fraction of FeO, namely ~64 wt%, after irradiation (S2-i). Given that the Fe cores (zero-valent Fe, $Fe^0$) remain about the same (S2-i) or grow (M1-D-i, M1-R-i M2-i), it seems reasonable to suspect that the region near the core would preferentially be $Fe^{2+}$ (i.e., FeO or $(Fe^{3+})[Fe^{2+},Fe^{3+}]O_4$), while the outer regions furthest from the core would be preferentially $Fe^{3+}$ (i.e., $(Fe^{3+})[Fe^{2+},Fe^{3+}]O_4$, $\gamma$-$Fe_2O_3$, or orthorhombic $Fe_2O_3$). Such a gradient in oxide stoichiometry has been observed in other sputtered core-shell iron oxides[51] and thin films,[52] and could be described as a partially oxidized magnetite. Using x-ray magnetic circular dichroism (XMCD), these authors showed that the outer surfaces of the iron oxide shells (with higher oxidation state than near the core interface) were preferentially spin-canted.[53]

It has been argued that a FeO ($Fe_{1-x}O$ or $Fe_{0.947}O$) phase should not be stable at all in 10 nm nanoparticles due to thermodynamic considerations of high surface energy.[54] Nonetheless, FeO domains can be shown to form kinetically as interim phases at interfaces from oxidation to hematite or reduction to magnetite.[43] It has been found that FeO can exist in three crystal structures, namely B1 (NaCl structure, $Fm\bar{3}m$), B2 (CsCl structure, $Pm\bar{3}m$), and B8 (NiAs, $P6_3mc$),[55] the stable phase being B1 which is what we observe. Similarly, an intermediate interface phase of maghemite $\gamma$-$Fe_2O_3$ can occur from oxidation of magnetite $Fe_3O_4$ to hematite $Fe_2O_3$.[43] Cation-deficient magnetite (i.e. partially oxidized or maghematized) is known to exist from geologic sediments as surface phases.[56] All this suggests that in fact gradations of oxygen stoichiometry are expected for iron oxide core-shell particles. Further detailed investigation of our irradiated particles with XMCD, Mössbauer spectroscopy, and electron energy loss spectroscopy (EELS) are needed to confirm the iron valence gradients suspected from GIXRD.

The precise mechanisms for reduction of the iron in the ion irradiated materials are still under investigation, but hypothetical mechanisms can be described. First, it is possible that oxygen interstitials could be introduced by primary knock-on displacement of atoms due to ion irradiation, and threshold displacement energies for Fe and O in iron oxide have been shown to be on the order of tens of eV,[57] sufficiently small for excitation by the 5.5 MeV Si+ ions. Subsequently, these oxygen atoms could thermally diffuse to the surface where they combine to form $O_2$ and release to vacuum during the ion irradiation. Alternatively, a process similar to photocatalysis could be taking place, where absorbed energy causes electron/ hole pair creation which enable reduction processes.[58]

### b. Change in Magnetic Behavior with Irradiation

Figure 5 shows the hysteresis loops, taken at 300 K, for all the films listed in Table 2 except for M1-R that was broken and unavailable for the measurement. Saturation magnetization, coercivity, and remanence values are listed in Table 4. For the core-shell NC/Si sample (S2), the irradiation-induced reduction in saturation might be expected since the shell changes from a ferrimagnetic material ($Fe_3O_4$)[59] to an antiferromagnetic material (FeO). Note that for bulk FeO the Néel temperature ($T_N$) is 198 K, but it has been shown that thin FeO layers with Fe on either side show $T_N$ well above room temperature due to exchange interactions with FM spins.[60] Given the large fractional decrease in ferromagnetic components, it is unexpected that the saturation magnetization did not decrease more, since FeO fraction is 64 wt%. It may be that Fe/FeO interfaces are ferrimagnetic, resulting in more saturation than would be expected given a mixture of the pure phases. It is also unclear the role played by the seemingly disordered or amorphous phases as seen in TEM. Further investigation is required to fully understand this behavior.

NC films were deposited onto the PLD-grown $Fe_3O_4$ single-crystal films on MgO substrates, but they did not fully adhere to the $Fe_3O_4$ films in all places. Magnetic measurements of unirradiated samples were therefore normalized to the fractional coverage of the NC film. This fractional area was used in order to calculate the normalized magnetization (emu/cm$^3$) using the density of Fe$_3$O$_4$ (5.2 g/cm$^3$) and the known mass of the NC deposited. Of these samples, half of each was used for ion irradiation, and the pieces chosen had full NC coverage. For irradiated samples, which had non-negligible fractions of iron species other than Fe$_3$O$_4$, the volume was taken from the unirradiated case with the appropriate area fraction (which had again changed from the unirradiated measurement), the volume magnetization calculated (emu/cm$^3$), and then the effective density of the NC was estimated from mass densities of the individual phases weighted per the mass fractions determined by XRD. The resulting theoretical densities are 5.45 and 5.36 g/cm$^3$ for M1-D-i and M2-i, respectively. Note that these estimates do not take into account any porosity but are merely used to normalize the magnetization so as to compare all on emu/g basis. In all cases (except the Fe$_3$O$_4$ single phase NC which were superparamagnetic before irradiation[28]), ion irradiation produced a decrease in saturation magnetization. This is because there is creation of a substantial component of FeO (5-9 wt%) in the case of the NC on PLD films. However, with the other core-shell sample where there was a very large FeO component (64 wt%), the saturation magnetization decreased only slightly. There was no clear trend in coercivity change with irradiation, with some of the samples showing slightly increased coercivity (S2, M2) while others slightly decreased (M1-R).

We did not observe any differences, magnetic or structural, between the samples where NC were deposited with the underlying PLD Fe$_3$O$_4$ film in the demagnetized versus the remnant state, neither in the unirradiated nor in the irradiated conditions. Additionally, we did not see any clear differences in the irradiated or unirradiated behaviors of the NC on the thin PLD Fe$_3$O$_4$ layer versus the thick one. We saw no evidence of biased alignment of the particles or preferential remnant magnetization as was previously observed in irradiated pure Fe$_3$O$_4$ NC.[28] The crystalline phase distributions were similar in all the MgO substrate samples within the groups of irradiated versus unirradiated samples.

### c. First order reversal curves

Fig. 6 shows the FORC diagrams of four NC films on Si, two unirradiated core/shell films with different core sizes (#2 and #4), and two irradiated films: the pure magnetite film (S1) and the core/shell film (S2). The irradiated core/shell film (Fig. 6c) consists of a very low coercivity component (~50 Oe) with a small spread in $H_b$, indicating little bias field or dipole interaction.[61] This demonstrates that the Fe cores are behaving nearly independently of one another, with the FeO shell (produced by irradiation) preventing interaction between them. The low coercivity is a size effect of the small Fe cores (~ 7 nm by XRD). In similar samples,[28,36] a distribution of particle sizes is seen, with some below the superparamagnetic (SPM) threshold. Only those particles above this size threshold will contribute to the coercivity, and the small distribution of $H_c$ seen in FORC is a consequence of size distribution. Though anisotropy suggests that the maximum threshold size for room temperature SPM in Fe should be 13 nm, these particles are much smaller, yet show coercivity at room temperature. This can be explained as being due to the influence of the oxide shell stabilizing the core moments, as seen also in Co-CoO particles.[62] Additionally, the oxide shell is likely ferri-magnetic, which makes the effective core larger. A comparison of similar core/shell samples with similar size (Fig. 6a) and larger cores (Fig. 6b) (~7 nm and 10 nm cores, respectively) show that with increasing core size the interactions increase (maximum on bias axis $H_b$ increases) as previously described for cluster glass systems.[33] This suggests that the nature of the cores of the irradiated film, which is surrounded by FeO, produces a shielding or diluting effect on the Fe cores which are otherwise similar in size to that shown in Fig. 6a.

On the other hand, the irradiated magnetite films show a broad coercivity distribution centered at ~400 Oe with a significant spread in $H_b$ due to particle interactions (Fig. 6d). Core-shell films before irradiation show similar centroids for the coercivity (400 – 500 Oe), but considerably smaller interactions. This could be due in part to the substantial increase in effective density in the irradiated films,[28,29] resulting in more dipolar interactions in the magnetite films. Recall however that the irradiated core/shell film of S2-i showed very small interactions due to the thick AFM shell.

## 4. CONCLUSIONS

One, two, or three-phase iron/ iron-oxide nanocluster films have been grown on Si(100) or MgO(100)/Fe$_3$O$_4$(100). Films exhibit an agglomerated structure at various size ranges according to transmission electron microscopy and helium ion microscopy. When irradiated to $10^{16}$ $Si^{2+}/cm^2$, all samples which contained a Fe core showed progressive reduction in Fe valence, with $Fe_2O_3/Fe_3O_4$ shells converting to $FeO/Fe_3O_4$ shells. Irradiated films also showed an overall increase in particle size, and a reduction in saturation magnetization. The irradiation-induced oxide phase change and magnetic behavior due to the antiferromagnetic shell is distinct from previously reported nanocluster systems of this type, and is likely due to oxygen evolution through defect sites created by the energetic ion beam.

## ACKNOWLEDGMENTS

This study was supported by the Pacific Northwest National Laboratory (PNNL) directed research & development (LDRD). The PNNL is operated for the U.S. Department of Energy by Battelle under Contract DE-AC05-76RL01830. Samples were prepared at the University of Idaho, supported by DOE under Contracts DE-FG02-07ER46386 and DE-FG02-04ER46142. A portion of the research was performed using the Environmental Molecular Sciences Laboratory (EMSL), a national scientific user facility sponsored by the DOE's Office of Biological and Environmental Research and located at PNNL. Work at UCD was supported by the NSF (DMR-1008791).

# REFERENCES

[1] X. X. Zhang, G. H. Wen, S. Huang, L. Dai, R. Gao, and Z. L. Wang, J. Magn. Magn. Mater. **231,** 9 (2001).
[2] Z. K. Wang, M. H. Kuok, S. C. Ng, D. J. Lockwood, M. G. Cottam, K. Nielsch, R. B. Wehrspohn, and U. Gösele, Phys. Rev. Lett. **89,** 027201 (2002).
[3] J. I. Martin, J. Nogues, K. Liu, J. L. Vicent, and I. K. Schuller, J. Magn. Magn. Mater. **256,** 449 (2003).
[4] J. Park, E. Kang, S. U. Son, H. M. Park, M. K. Lee, J. Kim, K. W. Kim, H. J. Noh, J. H. Park, C. J. Bae, J. G. Park, and T. Hyeon, Adv. Mater. **17,** 429 (2005).
[5] C. Boeglin, E. Beaurepaire, V. Halte, V. Lopez-Flores, C. Stamm, N. Pontius, H. A. Durr, and J. Y. Bigot, Nature **465,** 458 (2010).
[6] A. S. Edelstein and R. C. Cammarata, *Nanomaterials: Synthesis, Properties and Applications* (Taylor & Francis Group, New York:, 1996).
[7] Y. Zhao, C. Ni, D. Kruczynski, X. Zhang, and J. Q. Xiao, J. Phys. Chem. B **108,** 3691 (2004).
[8] H. Zeng, S. Sun, J. Li, Z. L. Wang, and J. P. Liu, Appl. Phys. Lett. **85,** 792 (2004).
[9] A. Sharma, Y. Qiang, D. Meyer, R. Souza, A. McConnaughoy, L. Muldoon, and D. Baer, J. Appl. Phys. **103,** 07A308 (2008).
[10] S. J. Cho, J. C. Idrobo, J. Olamit, K. Liu, N. D. Browning, and S. M. Kauzlarich, Chem. Mater. **17,** 3181 (2005).
[11] L. Zhou, J. Yuan, and Y. Wei, J. Mater. Chem. **21,** 2823 (2011).
[12] R. M. Wong, D. A. Gilbert, K. Liu, and A. Y. Louie, ACS Nano **6,** 3461 (2012).
[13] S. Talapatra, P. G. Ganesan, T. Kim, R. Vajtai, M. Huang, M. Shima, G. Ramanath, D. Srivastava, S. C. Deevi, and P. M. Ajayan, Phys. Rev. Lett. **95,** 097201 (2005).
[14] P. K. Kulriya, B. R. Mehta, D. K. Avasthi, D. C. Agarwal, P. Thakur, N. B. Brookes, A. K. Chawla, and R. Chandra, Appl. Phys. Lett. **96,** 053103 (2010).
[15] J. P. Nozières, M. Ghidini, N. M. Dempsey, B. Gervais, D. Givord, G. Suran, and J. M. D. Coey, Nucl. Instr. Meth. B **146,** 250 (1998).
[16] J. Ferré, C. Chappert, H. Bernas, J. P. Jamet, P. Meyer, O. Kaitasov, S. Lemerle, V. Mathet, F. Rousseaux, and H. Launois, J. Magn. Magn. Mater. **198–199,** 191 (1999).
[17] J. Q. Xiao, K. Liu, C. L. Chien, L. F. Schelp, and J. E. Schmidt, J. Appl. Phys. **76,** 6081 (1994).
[18] C. D'Orléans, J. P. Stoquert, C. Estournès, J. J. Grob, D. Muller, J. L. Guille, M. Richard-Plouet, C. Cerruti, and F. Haas, Nucl. Instr. Meth. B **216,** 372 (2004).
[19] L. G. Jacobsohn, J. D. Thompson, Y. Wang, A. Misra, R. K. Schulze, and M. Nastasi, Nucl. Instr. Meth. B **250,** 201 (2006).
[20] H. Kumar, S. Ghosh, D. K. Avasthi, D. Kabiraj, A. Mücklich, S. Zhou, H. Schmidt, and J.-P. Stoquert, Nanosc. Res. Lett. **6,** 155 (2011).
[21] C. Gavade, N. L. Singh, D. K. Avasthi, and A. Banerjee, Nucl. Instr. Meth. B **268,** 3127 (2010).
[22] A. Ishaq, A. R. Sobia, and L. Yan, J. Experim. Nanosc. **5,** 213 (2010).
[23] H. P. Xu, Y. Mao, J. Wang, B. Y. Xie, J. K. Jin, J. Z. Sun, W. Z. Yuan, A. Qin, M. Wang, and B. Z. Tang, J. Phys. Chem. C **113,** 14623 (2009).
[24] P. Esquinazi, D. Spemann, R. Höhne, A. Setzer, K. H. Han, and T. Butz, Phys. Rev. Lett. **91,** 227201 (2003).
[25] N. Nita, R. Schaeublin, and M. Victoria, J. Nucl. Mater. **329–333, Part B,** 953 (2004).
[26] C. M. Wang, D. R. Baer, J. E. Amonette, M. H. Engelhard, J. J. Antony, and Y. Qiang, Ultramicros. **108,** 43 (2007).
[27] J. A. Sundararajan, D. T. Zhang, Y. Qiang, W. Jiang, and J. S. McCloy, J. Appl. Phys. **109,** 07E324 (2011).
[28] W. Jiang, J. S. McCloy, A. S. Lea, J. A. Sundararajan, Q. Yao, and Y. Qiang, Phys. Rev. B **83,** 134435 (2011).
[29] J. McCloy, W. Jiang, J. Sundararajan, Y. Qiang, E. Burks, and K. Liu, AIP conf. proc. **1525,** 659 (2013).

[30]Y. Qiang, J. Antony, A. Sharma, J. Nutting, D. Sikes, and D. Meyer, J. Nanopart. Res. **8,** 489 (2006).
[31]M. Kaur, J. S. McCloy, and Y. Qiang, J. Appl. Phys. **113,** 17D715 (2013).
[32]J. E. Davies, O. Hellwig, E. E. Fullerton, G. Denbeaux, J. B. Kortright, and K. Liu, Phys. Rev. B **70,** 224434 (2004).
[33]J. E. Davies, J. Wu, C. Leighton, and K. Liu, Phys. Rev. B **72,** 134419 (2005).
[34]R. K. Dumas, K. Liu, C.-P. Li, I. V. Roshchin, and I. K. Schuller, Appl. Phys. Lett. **91,** 202501 (2007).
[35]C. R. Pike, A. P. Roberts, and K. L. Verosub, J. Appl. Phys. **85,** 6660 (1999).
[36]M. Kaur, J. S. McCloy, W. Jiang, Q. Yao, and Y. Qiang, J. Phys. Chem. C **116,** 12875 (2012).
[37]X. Kou, X. Fan, R. K. Dumas, Q. Lu, Y. Zhang, H. Zhu, X. Zhang, K. Liu, and J. Q. Xiao, Adv. Mater. **23,** 1393 (2011).
[38]T. C. Droubay, C. I. Pearce, E. S. Ilton, M. H. Engelhard, W. Jiang, S. M. Heald, E. Arenholz, V. Shutthanandan, and K. M. Rosso, Phys. Rev. B **84,** 125443 (2011).
[39]J. F. Ziegler, J. P. Biersack, and U. Littmark, *The Stopping and Range of Ions in Solids; available at* http://www.srim.org (Pergamon, New York, 1985).
[40]L. Machala, J. i. Tuček, and R. Zbořil, Chem. Mater. **23,** 3255 (2011).
[41]S. Ono, T. Kikegawa, and Y. Ohishi, J. Phys. Chem. Solids **65,** 1527 (2004).
[42]S. Ono and Y. Ohishi, J. Phys. Chem. Solids **66,** 1714 (2005).
[43]G. Ketteler, W. Weiss, W. Ranke, and R. Schlogl, Phys. Chem. Chem. Phys. **3,** 1114 (2001).
[44]F. Lu, J. Zhang, M. Huang, F. Namavar, R. C. Ewing, and J. Lian, J. Phys. Chem. C **115,** 7193 (2011).
[45]L. Jie, J. Zhang, F. Namavar, Y. Zhang, F. Lu, H. Haider, K. Garvin, W. J. Weber, and R. C. Ewing, Nanotechn. **20,** 245303 (2009).
[46]Y. Zhang, W. Jiang, C. Wang, F. Namavar, P. D. Edmondson, Z. Zhu, F. Gao, J. Lian, and W. J. Weber, Phys. Rev. B **82,** 184105 (2010).
[47]T. Mohanty, S. Dhounsi, P. Kumar, A. Tripathi, and D. Kanjilal, Surf. Coat. Techn. **203,** 2410 (2009).
[48]W. Jiang, L. Jiao, and H. Wang, J. Amer. Ceram. Soc. **94,** 4127 (2011).
[49]J. Antony, Y. Qiang, D. R. Baer, and C. Wang, J. Nanosci. Nanotechnol. **6,** 568 (2006).
[50]C. M. Wang, D. R. Baer, J. E. Amonette, M. H. Engelhard, Y. Qiang, and J. Antony, Nanotech. **18,** 255603 (2007).
[51]K. Fauth, E. Goering, G. Schutz, and L. T. Kuhn, J. Appl. Phys. **96,** 399 (2004).
[52]J. Korecki, B. Handke, N. Spiridis, T. Ślęzak, I. Flis-Kabulska, and J. Haber, Thin Solid Films **412,** 14 (2002).
[53]K. Fauth, E. Goering, and L. Theil-Kuhn, Mod. Phys. Lett. B **21,** 1197 (2007).
[54]A. Navrotsky, C. Ma, K. Lilova, and N. Birkner, Science **330,** 199 (2010).
[55]R. A. Fischer, A. J. Campbell, G. A. Shofner, O. T. Lord, P. Dera, and V. B. Prakapenka, Earth Plan. Sci. Lett. **304,** 496 (2011).
[56]A. V. Smirnov and J. A. Tarduno, J. Geophys. Res. **105,** 16457 (2000).
[57]G. S. Was, *Fundamentals of Radiation Materials Science* (Springer, New York, 2007).
[58]E. Casbeer, V. K. Sharma, and X.-Z. Li, Sep. Purific. Techn. **87,** 1 (2012).
[59]K. Liu, L. Zhao, P. Klavins, F. E. Osterloh, and H. Hiramatsu, J. Appl. Phys. **93,** 7951 (2003).
[60]S. Couet, K. Schlage, R. Rüffer, S. Stankov, T. Diederich, B. Laenens, and R. Röhlsberger, Phys. Rev. Lett. **103,** 097201 (2009).
[61]S. J. Cho, A. M. Shahin, G. J. Long, J. E. Davies, K. Liu, F. Grandjean, and S. M. Kauzlarich, Chem. Mater. **18,** 960 (2006).
[62]J. Nogués, V. Skumryev, J. Sort, S. Stoyanov, and D. Givord, Phys. Rev. Lett. **97,** 157203 (2006).
[63]H. E. Swanson and E. al., Nat. Bur. Stand. **539,** 3 (1955).
[64]R. Collongues and G. Chaudron, C. R. Hebd. Seances Acad. Sci **231,** 143 (1950).
[65]H. E. Swanson, H. F. McMurdie, M. C. Morris, and E. H. Evans, Nat. Bur. Stand. **25,** 31 (1967).

**TABLE 1.** Synthesis conditions for granular nanocluster films

| Sample name | Oxygen introduction point | Power (W) | Duration (min) | P, input Aggr Chamb (Torr) | P, output Aggr Chamb (Torr) | P, Dep Chamb (Torr) | Ar flow (sccm) | He flow (sccm) | $O_2$ flow (sccm) |
|---|---|---|---|---|---|---|---|---|---|
| Magnetite (S1)[28] | aggregation chamber | 200 | 30 | 1.1 | $8.8 \times 10^{-3}$ | $10^{-4}$ | 350 | 50 | 5 |
| Core/shell (S2)[29] | reaction chamber | 200 | 30 | 1.1 | $2 \times 10^{-4}$ | $10^{-4}$ | 350 | 50 | 3 |
| Mixed shell (M1-D) | reaction chamber | 200 | 60 | 1.3 | $10^{-4}$ | $10^{-4}$ | 350 | 50 | 3 |
| Mixed shell (M1-R) | reaction chamber | 200 | 60 | 1.3 | $10^{-4}$ | $10^{-4}$ | 350 | 50 | 3 |
| Mixed shell (M2) | reaction chamber | 200 | 60 | 1.3 | $10^{-4}$ | $10^{-4}$ | 350 | 50 | 3 |
| Core/shell (#4)[36] | reaction chamber | 200 | 45 | 2.5 | $10^{-4}$ | $10^{-4}$ | 600 | 50 | 2 |
| Core/shell (#2)[36] | reaction chamber | 200 | 45 | 1.5 | $10^{-4}$ | $10^{-4}$ | 400 | 50 | 2 |

Note: aggregation distance was constant at 310 mm

**TABLE 2.** Crystalline phases and average crystallite sizes for granular films before and after irradiation. Parentheses indicate standard deviation in the last digit. Note that the irradiated values marked with asterisk (*) are the crystallite sizes, and given the TEM results should be understood to represent "matrix" crystallite size and not necessarily shell thickness, which they do for the unirradiated case.

| Sample name | Oxygen introduction point | NC areal density (mg/cm$^2$) | NC film thickness (μm) | NC phase | Size (nm) | Wt% | NC phase | Size (nm) | Wt% |
|---|---|---|---|---|---|---|---|---|---|
| | | before irradiation | | Before irradiation | | | After irradiation | | |
| Magnetite (S1)[28] | aggregation chamber | 1.075 | - | $Fe_3O_4$ | 3(1) | 100 | $Fe_3O_4$ | 23(1) | 100 |
| Core/shell (S2)[29] | reaction chamber | - | - | Fe | 7.8(3) | 80.9 | Fe | 6.8(3) | 36.2 |
| | | | | | | | FeO | 12.5(9)* | 63.8 |
| | | | | $Fe_3O_4$ | 1.9(1) | 19.1 | | | |
| Core/shell (M1-D) | reaction chamber | 1.314 | 4.6 (HIM) | Fe | 9.0(5) | 2.6 | Fe | 29(1) | 7.8 |
| | | | | | | | FeO | 26(2)* | 8.5 |
| | | | | $Fe_3O_4$ | 1.9(1) | 73.9 | $Fe_3O_4$ | 17.9(8)* | 83.7 |
| | | | | $Fe_2O_3$ | 5(1) | 23.5 | | | |
| Core/shell (M1-R) | reaction chamber | 1.436 | 6.3 (HIM) | Fe | 9.0(5) | 2.6 | Fe | 33(1) | 8.5 |
| | | | | | | | FeO | 24(2)* | 4.1 |
| | | | | $Fe_3O_4$ | 1.9(1) | 76.7 | $Fe_3O_4$ | 15.2(4)* | 87.4 |
| | | | | $Fe_2O_3$ | 9(3) | 19.7 | | | |
| Core/shell (M2) | reaction chamber | 3.222 | 5.3 (HIM) | Fe | 8.8(5) | 2.1 | Fe | 31(1) | 5.4 |
| | | | | | | | FeO | 24(1)* | 5.0 |
| | | | | $Fe_3O_4$ | 1.8(1) | 78.4 | $Fe_3O_4$ | 16.8(4)* | 89.5 |
| | | | | $Fe_2O_3$ | 6(1) | 19.5 | | | |
| Core/shell (#4)[31] | reaction chamber | - | - | Fe | 10(1) | - | N/A | N/A | N/A |
| | | | | $Fe_3O_4$ | 2(1) | | | | |
| Core/shell (#2)[31] | reaction chamber | - | - | Fe | 7(1) | - | N/A | N/A | N/A |
| | | | | $Fe_3O_4$ | 2(1) | | | | |

N/A is not applicable; "-" indicates quantity was not measured.

**TABLE 3.** Details of the identified phases in the films before and after ion irradiation.

| Phase | Samples | PDF# | Symmetry | Space group | References |
|---|---|---|---|---|---|
| Fe | S1-u, S1-i, S2-u, M1-D-i, M1-R-i, M2-i | 006-0696 | Cubic | $Im\bar{3}m$ | 63 |
| FeO | S2-i, M1-D-i, M1-R-i, M2-i | 04-006-5424 | Cubic | $Fm\bar{3}m$ | 64 |
| $Fe_3O_4$ | All | 019-0629 | Cubic | $Fm\bar{3}m$ | 65 |
| $Fe_2O_3$ | M1-D-u, M1-R-u, M2-u | 056-1302 | Orthorhombic | Cmcm | 42 |

Note: -u = unirradiated; -i = irradiated

**TABLE 4.** Magnetic parameters at 300 K.

| Sample | Sample # | irradiation | Substrate | Core diam (nm) | Coercivity (Oe) | Saturation (emu/g) | Remanence (emu/g) | Reference |
|---|---|---|---|---|---|---|---|---|
| Magnetite | S1-u | No | Si | N/A | 0 | 4 | 0 | [28] |
| Magnetite | S1-i | Yes | Si | N/A | 250 | 44 | 9.4 | [28] |
| Core-shell | #4 | No | Si | 10 | 354 | 108 | 49 | [31] |
| Core-shell | #2 | No | Si | 7 | 102 | 90 | 16 | [31] |
| Core-shell | S2-u | No | Si | 8 | 45 | 89 | 14.5 | [29] |
| Core-shell | S2-i | Yes | Si | 7 | 56 | 77 | 16.5 | [29] |
| Core-shell on thin $Fe_3O_4$ PLD | M1-R-u | No | MgO | 9 | 500 | 63# | 18.8# | This work |
| Core-shell on thin $Fe_3O_4$ PLD | M1-R-i | Yes | MgO | 33 | 400 | 40# | 14.9# | This work |
| Core-shell on thick $Fe_3O_4$ PLD | M2-u | No | MgO | 9 | 395 | 45# | 13.1# | This work |
| Core-shell on thick $Fe_3O_4$ PLD | M2-i | Yes | MgO | 31 | 410 | 39# | 13.1# | This work |

Notes:
# Saturation is defined as the 10 kOe data point. The sample still appears to have a small paramagnetic component, which was not removed. The diamagnetic component of the MgO was not removed either, but it is a very small effect. Note that the saturation and the remanence are normalized to the mass of the NC film only, and the $Fe_3O_4$ underlying film is not treated here. It should be noted that the total moment of the PLD $Fe_3O_4$ underlying film was <10% of the total moment (emu) of the combined system with NC for the thick PLD film, and <1% of the total moment for the NC with thin PLD $Fe_3O_4$ film.

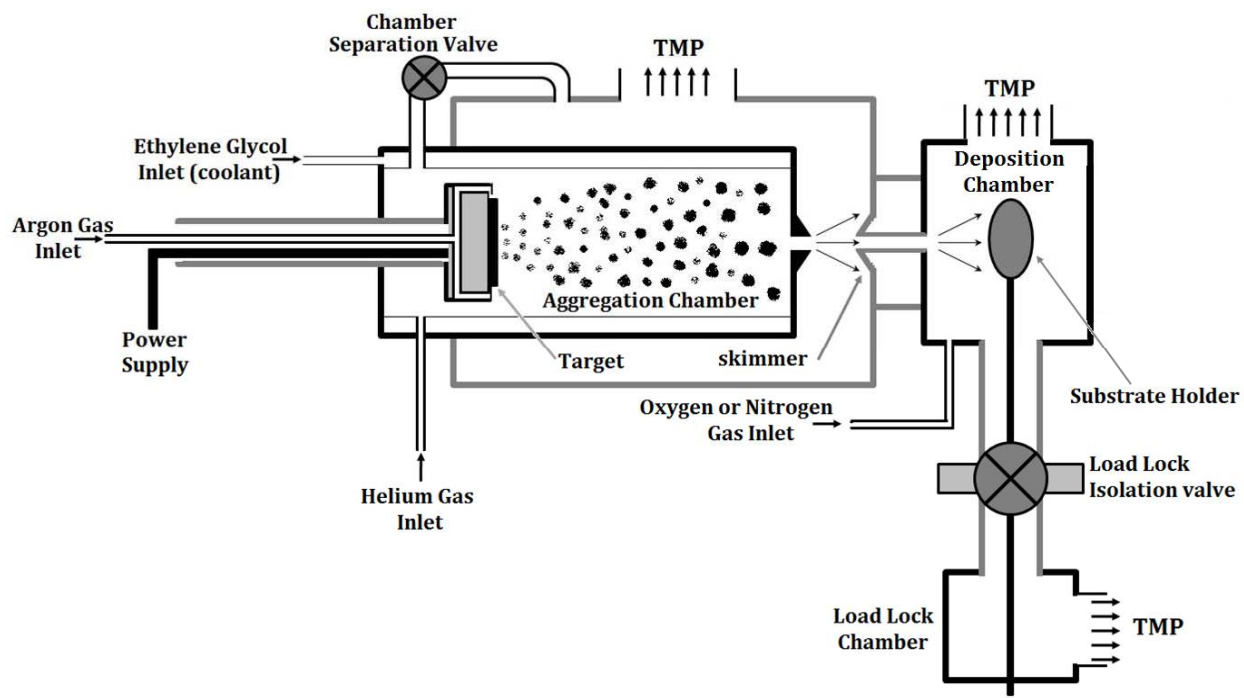

**FIGURE 1.** Schematic of the nanocluster deposition system. TMP indicates turbo molecular pump.

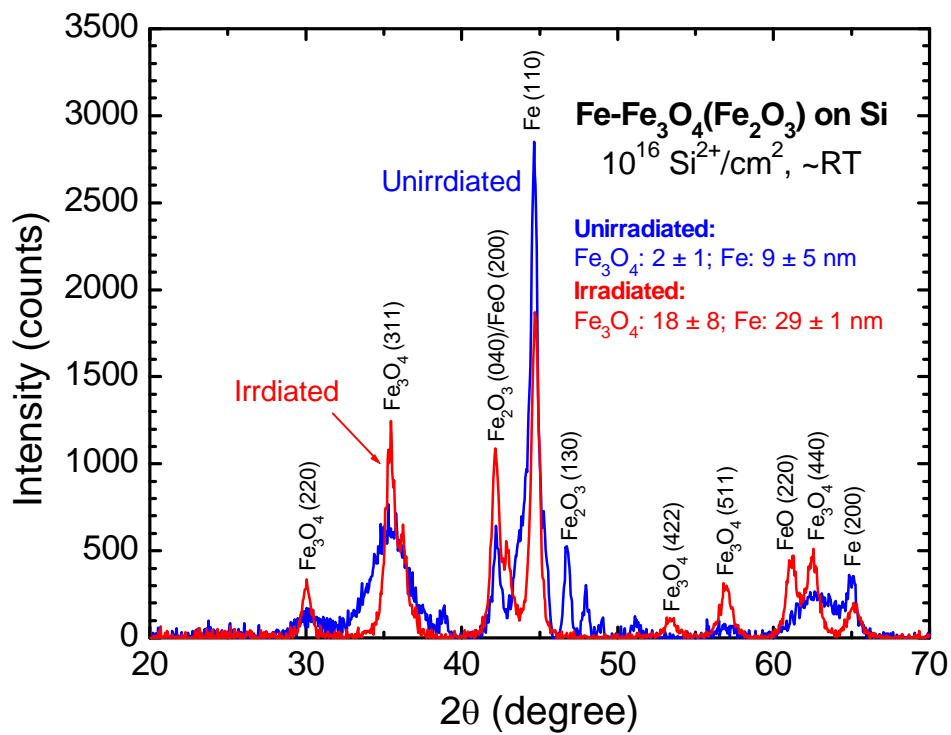

**FIGURE 2.** X-ray diffraction (XRD) spectrum of nanocluster film on MgO (M1-D) before and after irradiation. Both datasets are subtracted from their backgrounds. Inset text shows the crystallite sizes obtained from XRD line broadening.

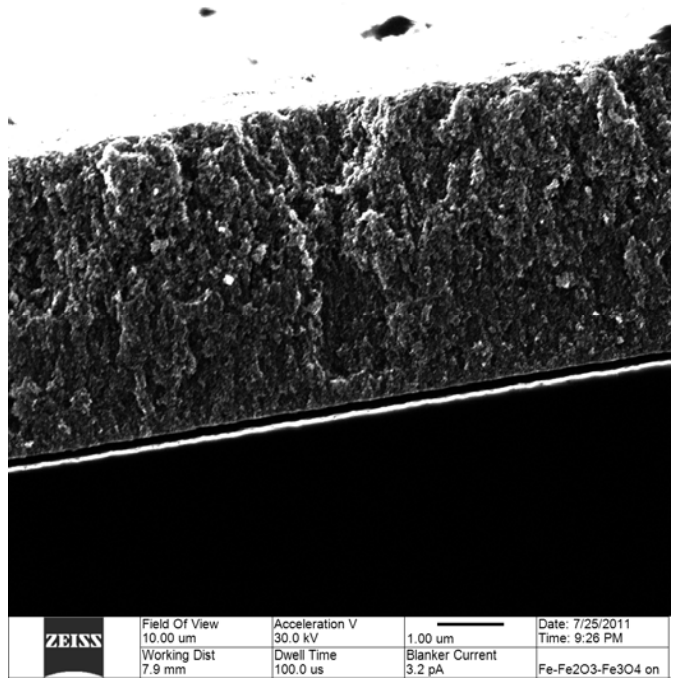

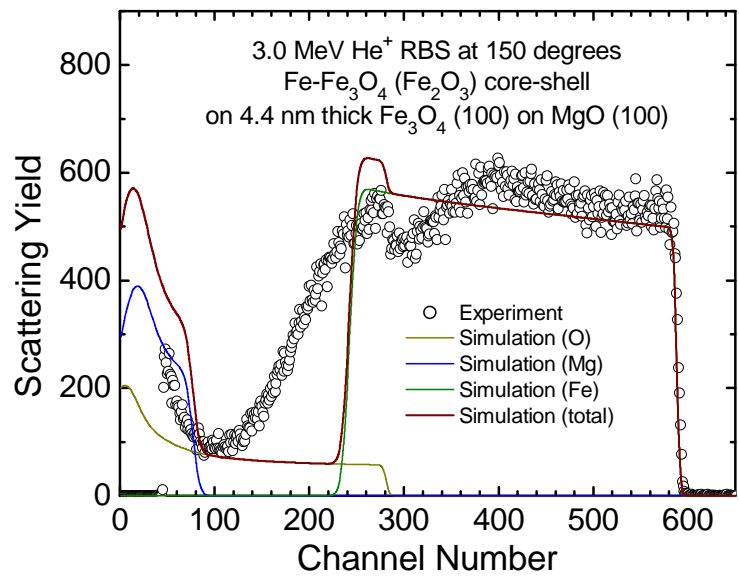

**FIGURE 3.** (a) Helium ion micrograph of sample M1-D cross-section prior to $Si^{2+}$ ion irradiation, and (b) 3.0 MeV $He^+$ Rutherford Backscattering Spectrometry (RBS) spectrum for the samples, together with simulation data (SIMNRA).

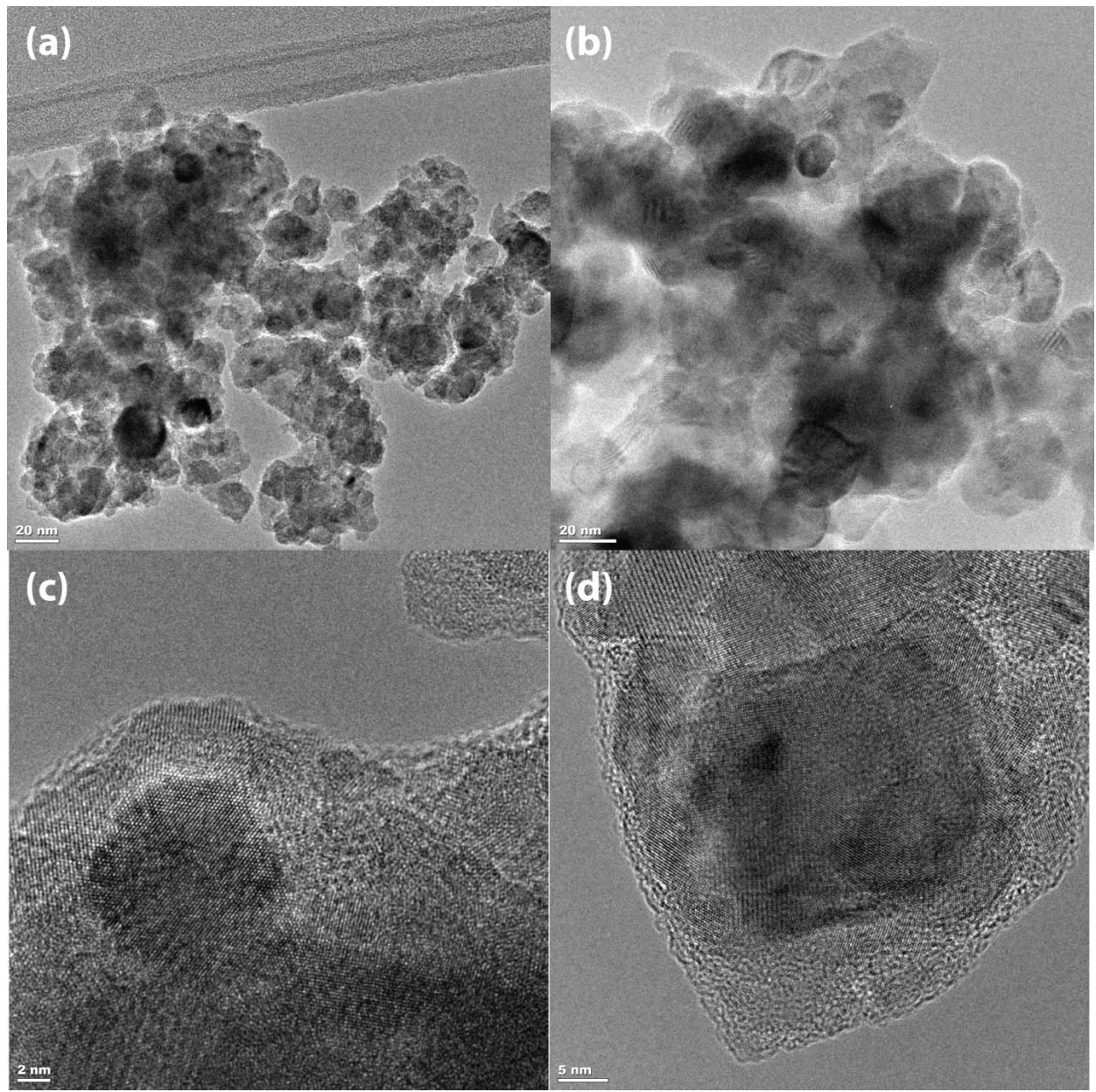

**FIGURE 4.** TEM micrographs of unirradiated (a,c) and ion irradiated (b,d) core-shell nanoparticles from sample M1-D. Fe core in (c) is ~12 nm, and in (d) is ~27 nm.

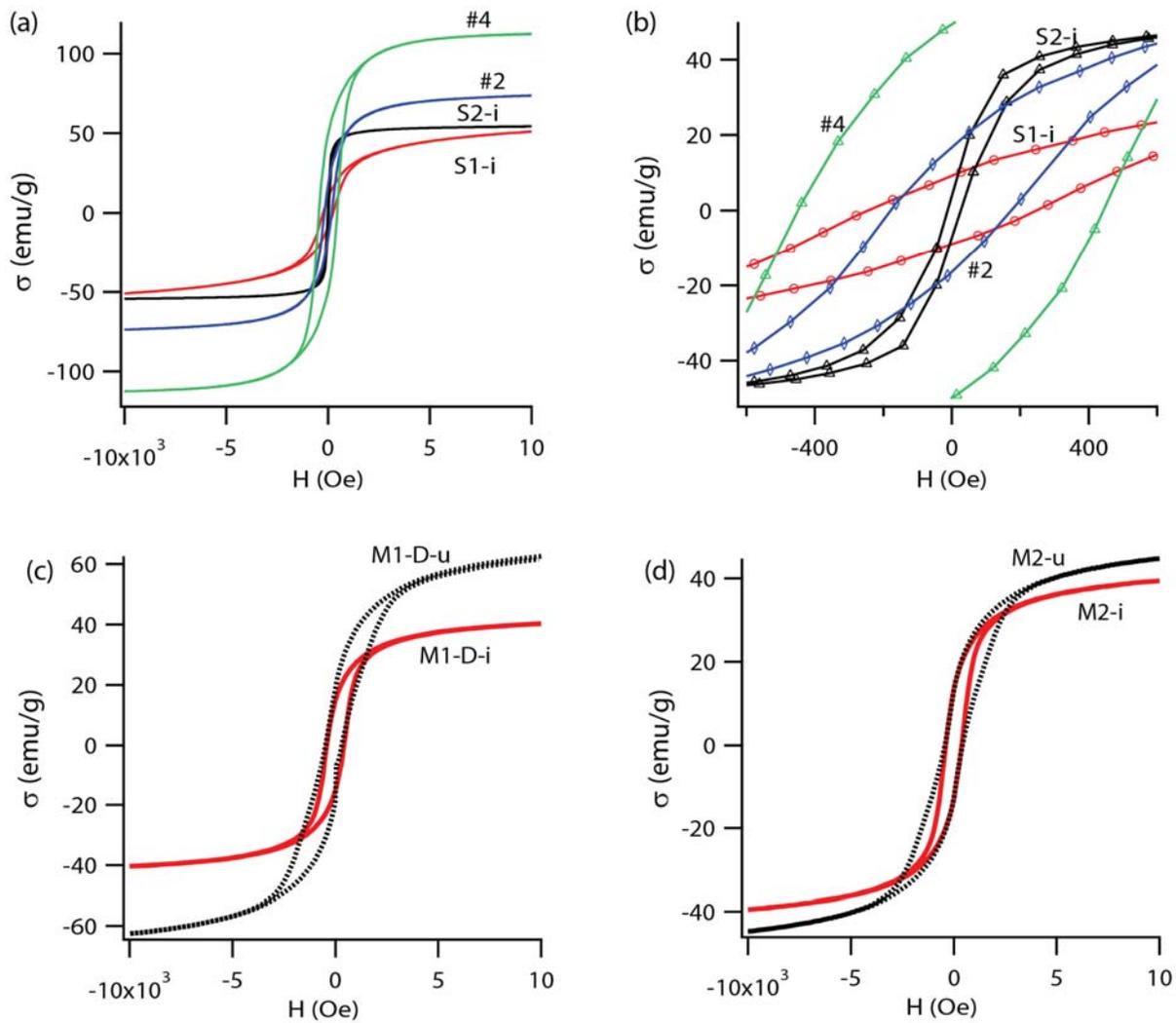

**FIGURE 5.** 300 K magnetization (a) nanoclusters on Si; (b) close-up of near zero field region of (a); (c) and (d): effect of irradiation on (c) M1-D and (b) M2, where NC mass normalized curve indicated for irradiated sample.

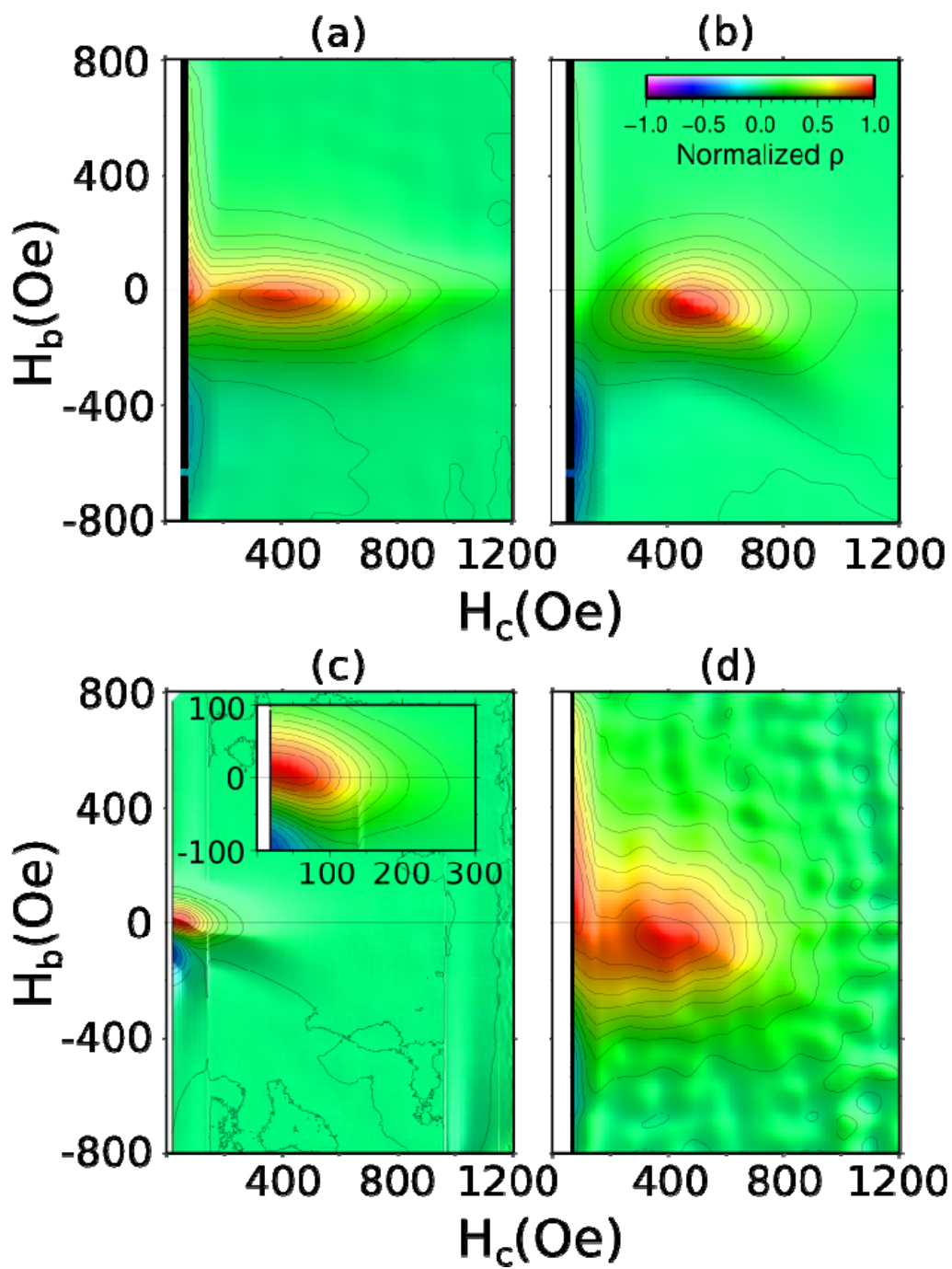

**FIGURE 6.** First order reversal curve diagrams (300 K). (a) unirradiated 7 nm core with 2 nm $Fe_3O_4$ shell (#2); (b) unirradiated 10 nm core with 2 nm $Fe_3O_4$ shell (#4); (c) irradiated 7 nm core with 12 nm FeO "shell" (S2-i); (d) irradiated single phase 23 nm $Fe_3O_4$ (S1-i).

25